\documentclass[intlimits,twoside,a4paper]{article}

\usepackage{amsmath,amssymb}
\usepackage{graphicx}
\usepackage{wrapfig}

\usepackage[T2A]{fontenc}
\usepackage[cp1251]{inputenc}

\usepackage[eqsecnum]{cmpj2}

\issue{2014}{17}{2}{23604}
\doinumber{10.5488/CMP.17.23604}

\title[Polymer concentration and chain length effects]%
{Self-consistent field theoretic simulations \\
of amphiphilic  triblock copolymer solutions: \\
Polymer concentration and chain length effects %
}

\author[X.-G. Han, Y.-H. Ma]{X.-G. Han\refaddr{label1,label2}\footnote{xghan0@163.com}\ ,
Y.-H. Ma\refaddr{label1,label2} }
\addresses{
\addr{label1} Inner Mongolia Key Laboratory for Utilization of Bayan
Obo Multi-Metallic Resources: \\ Elected State Key
Laboratory, Inmongolia Science and Technology University, Baotou
014010, China
\addr{label2} School of Mathematics, Physics and
Biology, Inmongolia Science and Technology University, \\ Baotou
014010, China}

\date{Received December 18, 2013, in final form March 28, 2014}
\authorcopyright{X.-G. Han, Y.-H. Ma, 2014}

\begin{document}

\maketitle

\begin{abstract}
Using the self-consistent field lattice model, polymer concentration
$\bar{\phi}_{{P}}$
 and chain length $N$ (keeping the length ratio of hydrophobic to hydrophilic blocks constant)
the effects on temperature-dependent behavior of micelles are studied, in
amphiphilic symmetric ABA triblock  copolymer solutions. When chain
length is increased, at fixed $\bar{\phi}_{{P}}$, micelles occur at
higher temperature. The variations of average volume fraction of
stickers $\bar{\phi}_\textrm{co}^\textrm{s}$ and the lattice site numbers
$N_\textrm{co}^\textrm{ls}$ at the micellar cores with temperature are dependent
on $N$ and $\bar{\phi}_{{P}}$, which demonstrates that the
aggregation of micelles depends on $N$ and $\bar{\phi}_{{P}}$.
Moreover, when $\bar{\phi}_{{P}}$ is increased, firstly a peak
appears on the curve of  specific heat $C_V$ for unimer-micelle
transition, and then in addition a primary peak, the secondary peak,
which results from the remicellization, is observed on the curve
of $C_V$. For a long chain, in intermediate and high concentration
regimes, the shape of specific heat peak markedly changes, and the
peak tends to be a more broad peak. Finally, the aggregation
behavior of micelles is explained by the aggregation way of
amphiphilic triblock copolymer. The obtained results are helpful in
understanding the micellar aggregation process.

\end{abstract}

\section{Introduction}

Polymeric micelles constitute a unique class of nanomaterials having a
typical core-shell morphology. They are formed from amphiphilic
block- or graft-copolymers in a selective solvent, where the
non-soluble parts self-assemble to form the core of the micelles and
the soluble parts form the solvated shell. The properties of micelle
can be changed by the solution conditions such as concentration,
temperature, and chain architecture. Such self-assembly phenomena of
amphiphilic molecules are of principal importance in many biological
and industrial processes. Recently, self-assembled bolaamphiphile
nanotubes have been used as templates to produce metal-coated
nanowires \cite{Mats2000}. A detailed understanding of the
aggregation process is crucial to understand and eventually control
their formation for the related applications of micelles.

The triblock copolymers, made up of poly(ethyleneoxide) (PEO) and
poly(propyleneoxide) (PPO) \linebreak blocks, which are experimentally studied
as amphiphilic molecules,  have been the subject of intense research
over the last two decades due to their unique solution
behavior \cite{Ries2003,Tron2008}. Furthermore, the arrangement of
the PPO and PEO blocks in the chain is the key factor affecting
self-aggregation and phase behavior of these copolymers, which are
well documented in literature
 \cite{Tron2008,Zhou1994,Chu1996,Wu2005,Erri2004,Pate2010}. The
temperature induced aggregation behavior of triblock copolymers in
aqueous solutions has received great attention during the recent decades due
to their fundamental and practical
importance \cite{Tron2008,Chu1996,Wu2005,Erri2004,Pate2010,Hugo2011}.
Compared with experimental studies, however, related theoretical
studies are few, especially to account for the effect of chain
architecture. Han et al.  \cite{Han2013} investigated the effects of
the length of each hydrophobic end block and polymer concentration
 on micellar aggregation in amphiphilic symmetric ABA
triblock copolymer solutions. It is found that the broadness of
transition is affected by the length of  hydrophobic end blocks
(i.e., chain length). However, in associative polymers \cite{Han2012},
when the total length of hydrophilic blocks is decreased, keeping
the chain length constant, the broadness of transition concerned
micelles increases. It is an obvious conclusion drawn that the
broadness of the transition  changes due to the length ratio of
hydrophobic to hydrophilic blocks. The chain length is an important
parameter to understand the thermodynamics of block copolymers in a
selective solvent. In amphiphilic triblock copolymer solutions,
however, the effect of chain length on micellar aggregation
behavior has not been clarified so far, the length ratio of hydrophobic
to hydrophilic blocks remaining constant.

As a mesoscopic polymer theory, the self-consistent field theory
(SCFT) has its origin from the field theoretical approach by
Edwards \cite{Edwa1965} and was explicitly adopted to deal with block
copolymer structures by Helfand \cite{Helf1975}. In recent years,
Matsen and Schick proposed a powerful numerical spectral method that
could be used to deal with complex
microphases \cite{Mats1994,Mats1995}. This method is accurate enough
but requests a prior knowledge of the symmetry of an ordered
structure, which has hindered its application in predicting
microphases of complex copolymer structures. Subsequently, Drolet
and Fredrickson suggested a new combinatorial screening
method \cite{Drol1999,Drol2001}, which involves a direct
implementation of SCFT in real space in an adaptive arbitrary cell.
This method proves to be very successful and can be applied to
complex copolymer melts. It has also been extended to predict the
nanostructures of polymer-grafted nanoparticles \cite{Reis2005},
which have potential applications in the design and synthesis of
hierarchical materials. In addition, SCFT allows us to investigate
the aggregate morphology of amphiphilic block copolymers and their
blends in a dilute solution \cite{He2004,Jian2005,Wang2005,Wang2011}.
Recently, Matsen extends SCFT to treat diblock copolymers with
nongaussian chain of low molecular weight \cite{Mats2012}.

A lattice model is introduced to self-consistent mean-field theory
to treat microphase separation for rod-coil block
copolymers \cite{Chen2006,Chen2007,Chen2008}. In our previous
papers \cite{Han2010,Han2011,Han2012,Han2013}, we have used the SCFT
lattice model to study the phase behavior of physically associating
polymer solutions. It is found  that chain architecture and
polymer concentration are important factors which affect the
property of temperature-dependent aggregation behavior. Now, in
amphiphilic ABA symmetric triblock copolymer solutions, we study
chain length and polymer concentration effects on aggregation
behavior. It is found that although the length ratio of
hydrophobic to hydrophilic blocks remains constant, the increase
in the aggregation degree of micelles is also dependent on the chain
length, and it is explained by the way of aggregation of amphiphilic
triblock copolymer.

\section{Theory\label{sec2}}

This section briefly describes the self-consistent field theory
(SCFT) lattice model for $n_{P}$ amphiphilic symmetric  ABA
triblock copolymers which are assumed to be incompressible. Each
block molecule consists of $N_\textrm{ns}$ nonsticker segments forming
the middle B block and $N_\textrm{st}$ sticker
segments forming each end A block, distributed over a lattice. At the same time, $%
n_{h}$ solvent molecules are placed on the vacant lattice sites.
Polymer monomers and solvent molecules have the same size and each
occupies one lattice site. The total number of lattice sites is $N_{{L}}$ $=$ $%
n_{h}+n_{P}N$. The transfer matrix $\lambda$ is used to describe
the polymer chain, which depends only on the chain model used. We
assume that
\begin{equation}
\lambda_{r_{s}-r'_{s-1}}^{\alpha _{s}-\alpha
_{s-1}}=
\left\{
  \begin{array}{ll}
    0\,, & \hbox{$\alpha _{s}=\alpha _{s-1}$\,,} \\
    {1}/(z-1)\,, & \hbox{{otherwise}\,.}
  \end{array}
\right.
%
\end{equation}
Here, $r'$ denotes the nearest
neighboring site of $r$. $r_{s}$ and $\alpha _{s}$ denote the position
and bond orientation of the $s$-th segment of the copolymer,
respectively. $\alpha$ can be any of the allowed bond orientations
depending on the lattice model used. $z$ is the coordination number
of the lattice. This means that the chain is described as a random
walk without a possibility of direct backfolding. Although
self-intersections of a chain are not permitted, the excluded volume
effect is sufficiently taken into account \cite{Medv2001}. $G^{\alpha
_{s}}(r,s|1)$ is the end segment distribution function of the $s$-th
segment of the chain. Following the scheme of Schentiens and
Leermakers \cite{Leer1988}, it is evaluated from the following
recursive relation:
\begin{equation} G^{\alpha _{s}}(r,s|1)=G(r,s)\sum_{r_{s-1}^{\prime
}}\sum_{\alpha _{s-1}}\lambda _{r_{s}-r_{s-1}^{\prime }}^{\alpha
_{s}-\alpha _{s-1}}G^{\alpha _{s-1}}(r^{\prime },s-1|1),
\label{free}
\end{equation} where $G(r,s)$ is the free segment
weighting factor and is expressed as
\[
G(r,s)=
\left\{
  \begin{array}{ll}
    \exp[-\omega_\textrm{st}(r_{{s}})]\,, & \hbox{$s\in \textrm{st}$,} \\
    \exp[-\omega_\textrm{ns}(r_{{s}})]\,, & \hbox{$s\in \textrm{ns}$.}
  \end{array}
\right.
\]
The initial condition is $G^{\alpha _{1}}(r,1|1)=G(r,1)$ for
all the values of $\alpha _{1}$. $\sum_{r_{s-1}^{\prime
}}\sum_{\alpha _{s-1}}$ means the summation over all the possible
positions and orientations of the $(s-1)$-th segment of the chain.
Another end segment distribution function $G^{\alpha _{s}}(r,s|N)$
is evaluated from the following recursive relation: \begin{equation}
G^{\alpha _{s}}(r,s|N)=G(r,s)\sum_{r_{s+1}^{\prime }}\sum_{\alpha
_{s+1}}\lambda _{r_{s+1}^{\prime }-r_{s}}^{\alpha _{s+1}-\alpha
_{s}}G^{\alpha _{s+1}}(r^{\prime },s+1|N),
\end{equation}
with the initial condition $G^{\alpha _{N}}(r,N|N)=G(r,N)$ for all
the values of $\alpha _{N}$.

 In this simulation, the free energy in the canonical ensemble $F$ is defined as

\begin{equation}
\frac{F[\omega _{+},\omega _{-}]}{k_\textrm{B}T}=\sum_{r}\left\{ \frac{1}{4\chi }%
\omega _{-}^{2}(r)-\omega _{+}(r)\right\} -n_{P}\ln Q_{{P}}[\omega
_\textrm{st},\omega _\textrm{ns}]-n_{h}\ln Q_{h}[\omega _{h}],  \label{free0}
\end{equation}%
where $\chi $ is the Flory-Huggins interaction parameter in the
solutions, which equals $\frac{z}{2k_\textrm{B}T}\epsilon$, $z$ is the
coordination number of the lattice used. $Q_{{h}}$ is the partition
function of a solvent molecule subjected to the field $\omega
_{h}(r)=$ $\omega _{{+}}(r)$, which is defined as
$Q_{{h}}=\frac{1}{n_{h}}\sum_{r}\exp\left[ -\ \omega
_{h}(r)\right]$. $Q_{{P}}$ is the partition function of a
noninteraction polymer chain subjected to the fields $\omega
_\textrm{st}(r)=\omega _{{+}}(r)-\omega _{{-}}(r)$ and $\omega
_\textrm{ns}(r)=\omega _{{+}}(r)$, which act on sticker and nonsticky
segments, respectively. $Q_{{P}}$ is expressed as $Q_{{_P}}=\frac{1%
}{N_{L}}\frac{1}{z}\sum_{r_{N}}\sum_{{\alpha }_{N}}G^{\alpha _{N}}(r,N|1)$, where $r_{N}$ and ${\alpha }_{N}$ denote the position and
orientation of the $N$-th segment of the chain, respectively.
$\sum_{r_{N}}\sum_{\alpha _{N}}$ means the summation over all the
possible positions and orientations of the $N$-th segment of the
chain. Minimization of the free energy function $F$ with $\omega
_{{-}}(r)$ and $\omega _{{+}}(r)$ leads to the following saddle
point equations:
\begin{equation}
\omega _{{-}}(r)=2\chi \phi _\textrm{st}(r),  \label{scf1}
\end{equation}%
\begin{equation}
\phi _\textrm{st}(r)+\phi _\textrm{ns}(r)+\phi _{h}(r)=1, \end{equation} where
\begin{equation}
\phi _\textrm{st}(r)=\frac{1}{N_{L}}\frac{1}{z}\frac{n_{P}}{Q_{P}}\sum_{s\in \textrm{st}%
}\sum_{\alpha _{s}}\frac{G^{\alpha _{s}}(r,s|1)G^{\alpha _{s}}(r,s|N)}{G(r,s)%
}
\end{equation}
and
\begin{equation}
\phi _\textrm{ns}(r)=\frac{1}{N_{L}}\frac{1}{z}\frac{n_{P}}{Q_{P}}\sum_{s\in \textrm{ns}%
}\sum_{\alpha _{s}}\frac{G^{\alpha _{s}}(r,s|1)G^{\alpha _{s}}(r,s|N)}{G(r,s)%
}
\end{equation}
are the average numbers of sticker and nonsticky segments at $r$,
respectively, and $\phi_{h}(r)=({1}/{N_{{L}}})({n_{h}}/{\
Q_{{_h}}})$ $\times\exp \left[ -\ \omega _{h}(r)\right]$ is the average
numbers of solvent molecules at $r$.

In our calculations, real space method is implemented to solve the
SCFT equations in a cubic lattice with periodic boundary conditions,
which is similar to our previous paper \cite{Han2010}. The
configuration from SCFT equations is taken as a saddle point configuration. By
comparing the free energies of the observed states from different
initial fields, a relative stability of the observed morphologies
can be obtained.

\section{Result and discussion\label{sec3}}

\begin{figure}
\centering
\includegraphics[width=0.55\textwidth]{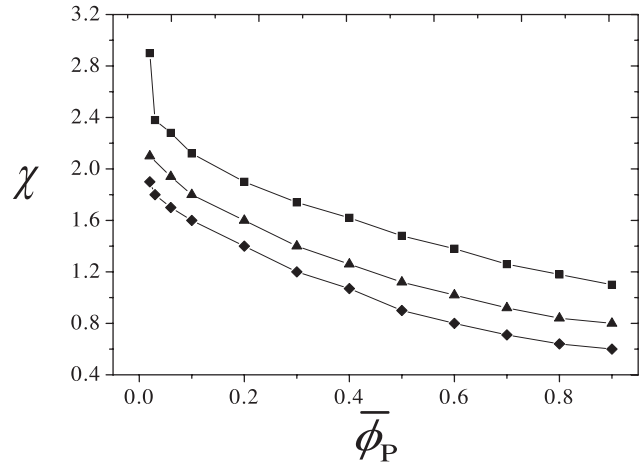}
\caption{The
phase diagram for the amphiphilic
 symmetric ABA tribolck copolymers with different chain length $N$.
The boundary between homogenous solutions and micelle morphology is
obtained. The squares, triangles and diamonds correspond to the
boundaries for $N=17, 26, 34$, respectively.\label{phadia}}
\end{figure}

In our studies, the property of symmetric ABA triblock copolymers is
characterized by three tunable molecular parameters: $\chi$ (The
Flory-Huggins interaction parameter), $N$ (The chain length of
copolymer) and $N_\textrm{st}/N_\textrm{ns}$ (the length ratio of each hydrophobic
end block to hydrophilic middle block). In this paper, when the chain
length is changed, the value of $N_\textrm{st}/N_\textrm{ns}(\simeq0.23)$ remains
constant. The aggregation behavior of micelle morphologies is
focused when the length of copolymer is changed. Figure~\ref{phadia}
shows the phase diagram of the systems with different chain length
$N$. When $\chi$ is increased, the unimer-micelle transition occurs.
At fixed $N$, the $\chi$ value on micellar boundary increases with
decreasing $\bar{\phi}_{P}$. When $N$ is increased, at fixed
$\bar{\phi}_{P}$, the $\chi$ value on micellar boundary shifts to
a small value. It is noted that although the length ratio of each
hydrophobic end block to hydrophilic middle block remains
constant, the increase in the chain length of copolymer is also
favorable to the occurrence of micelles in the system.

\begin{figure}[!t]
\centering
\qquad
\includegraphics[width=0.6\textwidth]{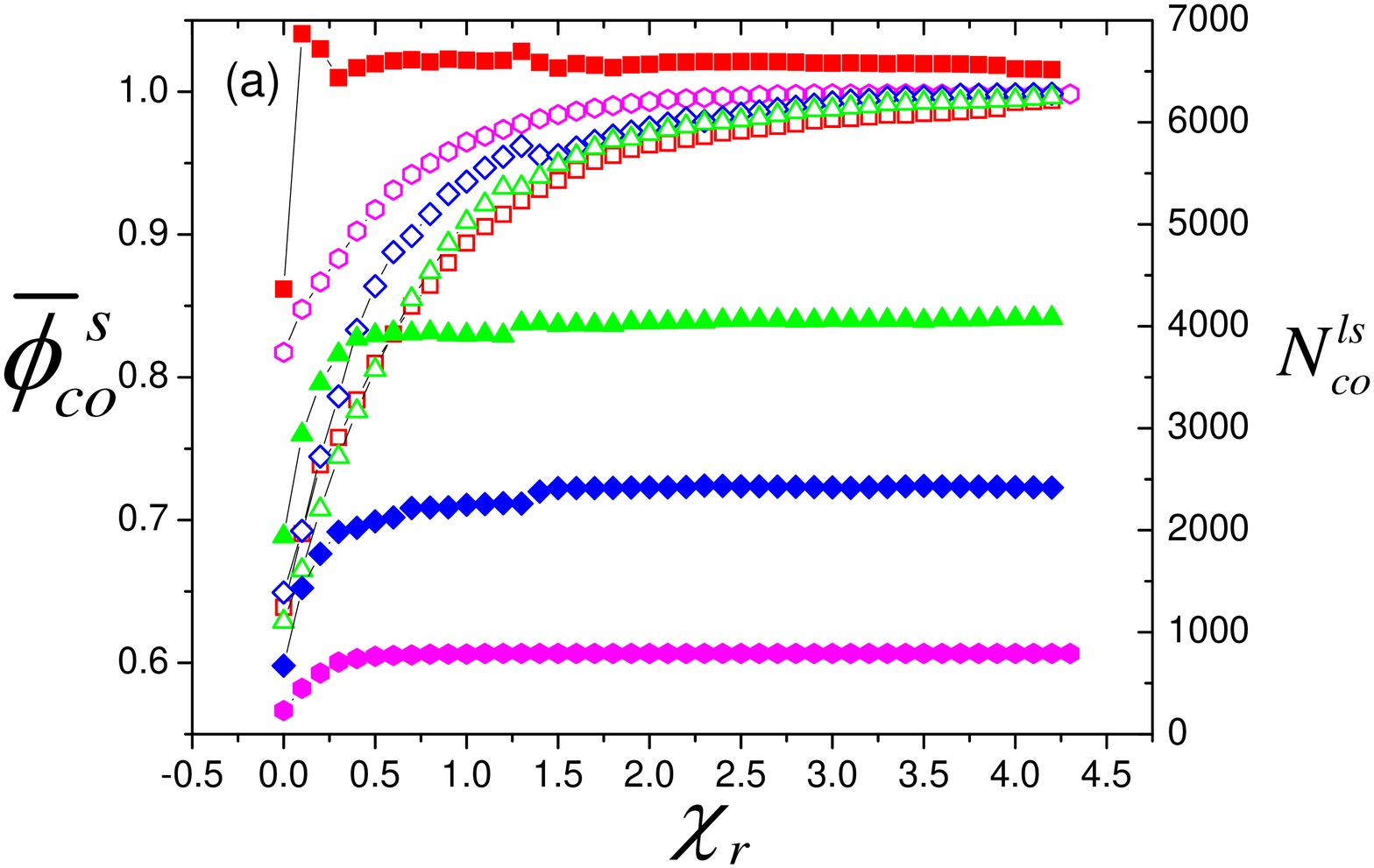}\\[-2ex]
\includegraphics[width=0.6\textwidth]{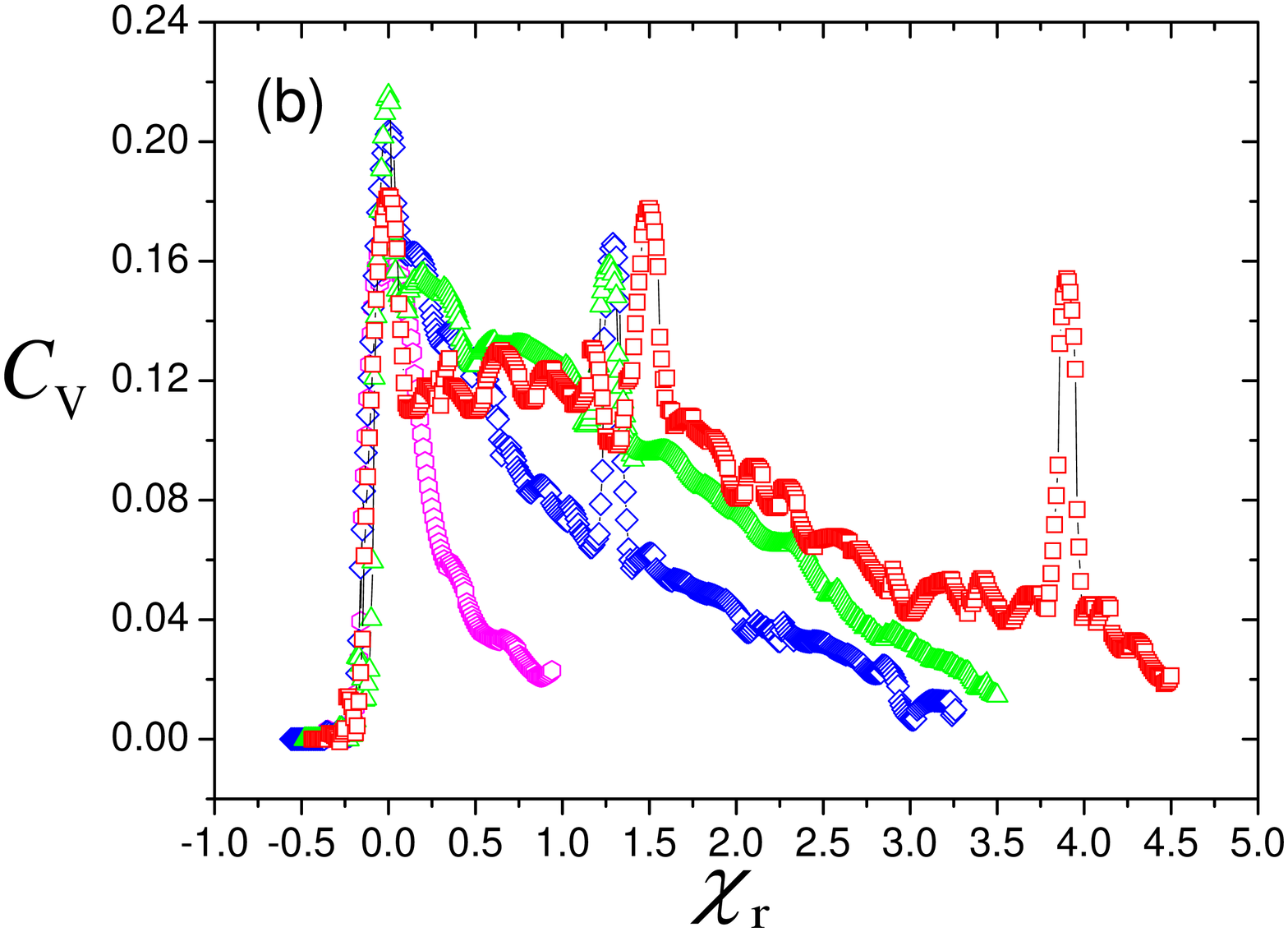}
\caption{(Color online) The variations of average volume fractions of stickers
$\bar{\phi} _\textrm{st}$ and lattice site numbers $N_\textrm{co}^\textrm{ls}$ at the
micellar cores in different amphiphilic
 ABA tribolck copolymers with the $\chi$ deviation from
micellar boundary $\chi_r$, for various $\bar{\phi}_{P}$ at $N=26$
is presented in figure~\ref{Cva}~(a). The open and solid, open and
solid triangles, open and solid diamonds, and open and solid
hexagons denote the $\bar{\phi} _\textrm{st}$ and $N_\textrm{co}^\textrm{ls}$ for
$\bar{\phi}_{P}=0.8, 0.5, 0.3, 0.1$, respectively; The changes of
heat capacity for different $\bar{\phi}_{P}$ in figure~\ref{Cva}~(a)
with $\chi_r$ are shown in figure~\ref{Cva}~(b). The squares,
triangles, diamonds and hexagons denote the case of
$\bar{\phi}_{P}=0.8, 0.5, 0.3, 0.1$, respectively. \label{Cva}}
\end{figure}

\begin{figure}[!t]
\vspace{-5mm}
\centering
\includegraphics[width=0.6\textwidth]{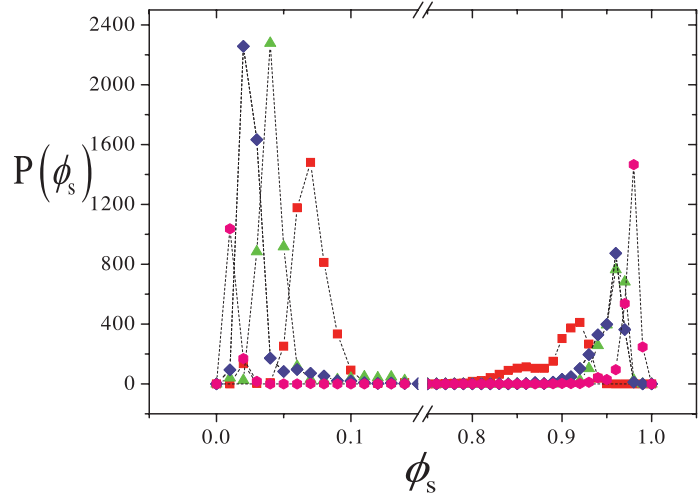}
\caption{(Color online) The distributions of the numbers of micellar core sites
with $\phi _{\mathrm{co}}^{\mathrm{s}}$  at $\bar{\phi}_{P}=0.3 $ and $N=26$. The squares,
triangles, diamonds and pentacles correspond to $\chi_r =0.7$, $1.3$,
$1.4$, $2.4$, respectively.  \label{dist}}
\end{figure}

\begin{figure}[!t]
\centering
\qquad\includegraphics[width=0.65\textwidth]{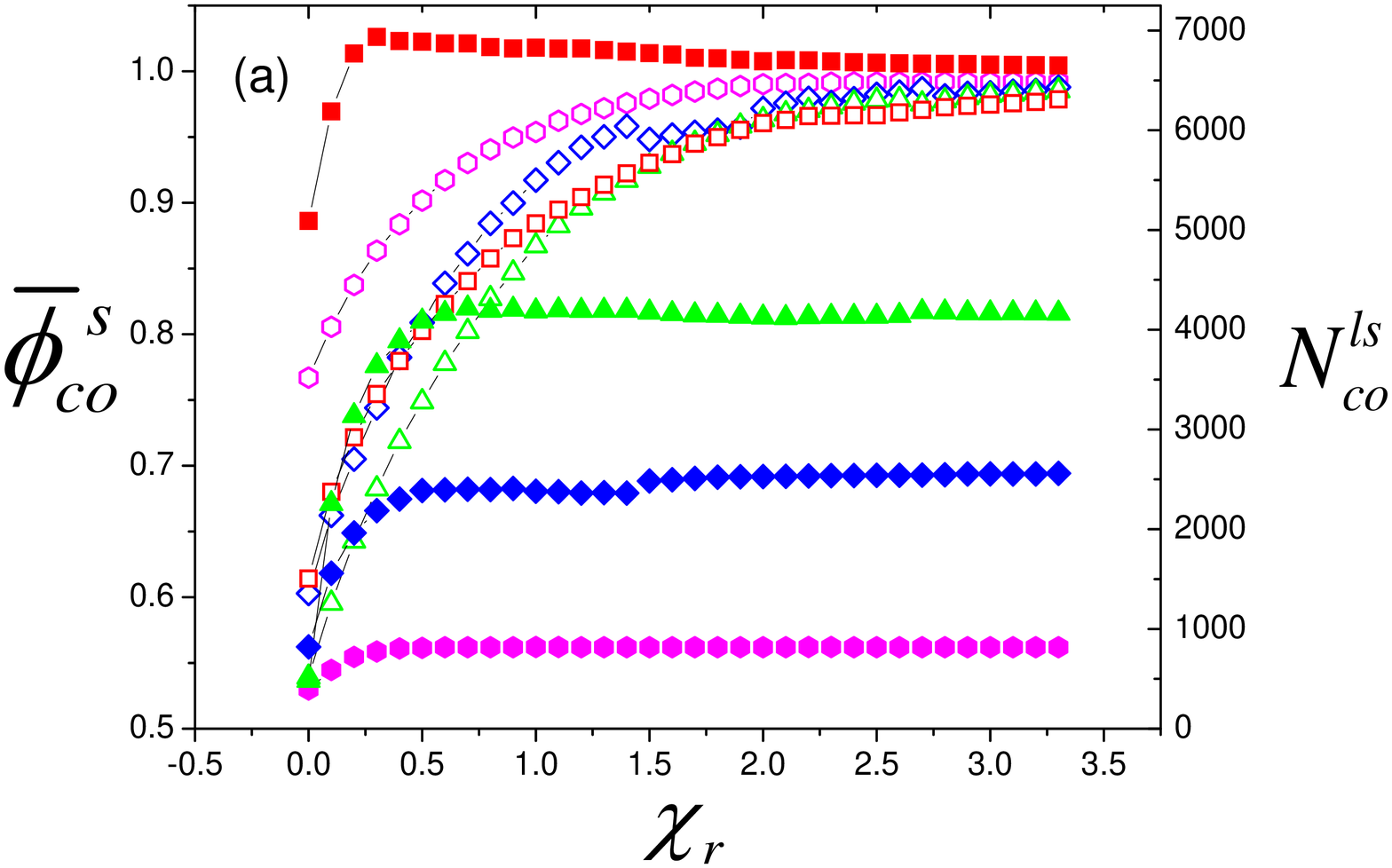}\\[-2ex]
\includegraphics[width=0.6\textwidth]{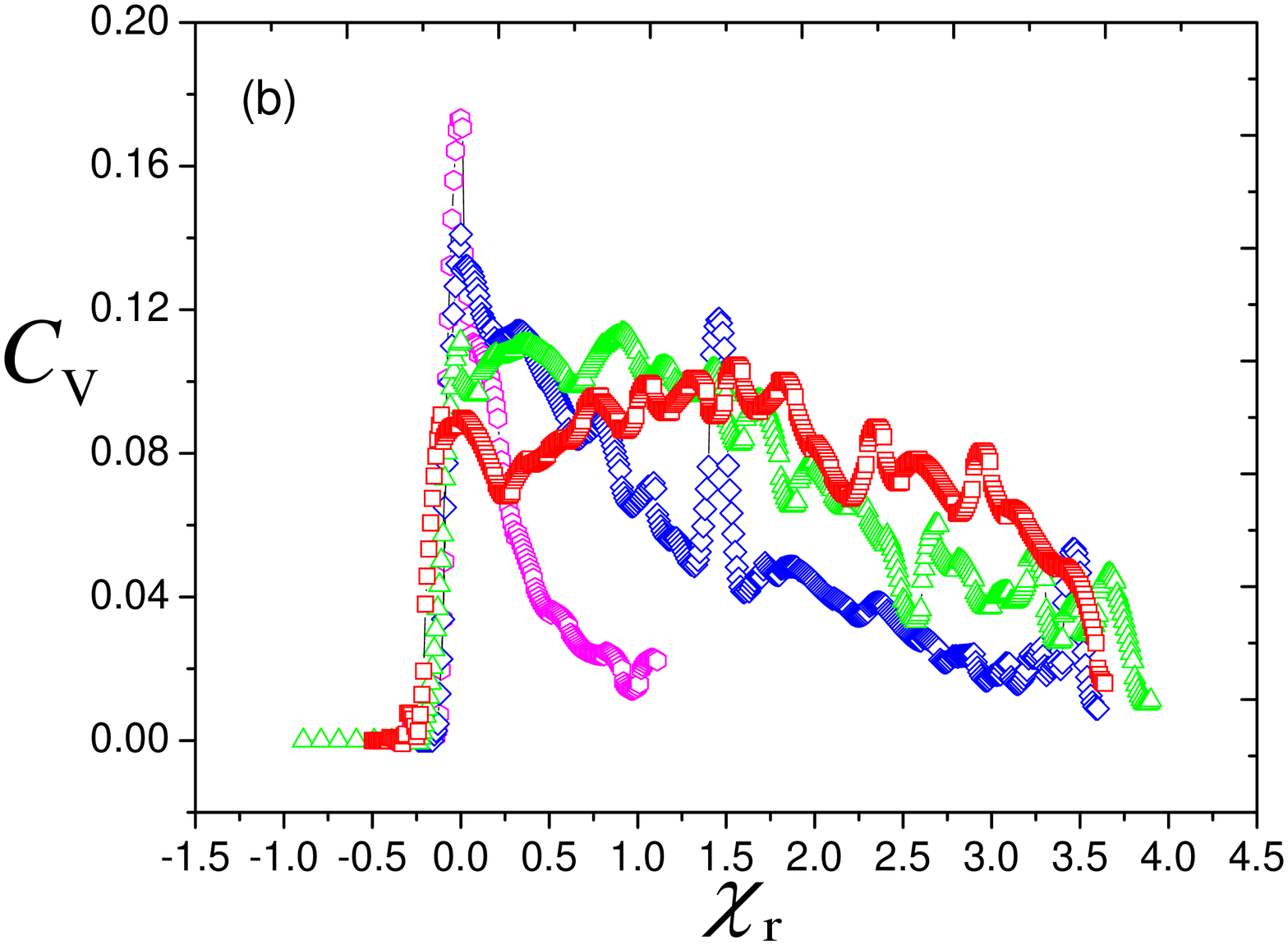}
\caption{(Color online) The variations of the average volume fractions of stickers
$\bar{\phi} _\textrm{st}$ and lattice site numbers $N_\textrm{co}^\textrm{ls}$ at the
micellar cores in different amphiphilic
 symmetric ABA tribolck copolymers with the $\chi$ deviation from
micellar boundary $\chi_r$, for various $\bar{\phi}_{P}$ at $N=34$
is presented in figure~\ref{Cvb}~(a). The open and solid squares,
open and solid triangles, open and solid diamonds, and open and
solid hexagons denote the $\bar{\phi} _\textrm{st}$ and $N_\textrm{co}^\textrm{ls}$ for
$\bar{\phi}_{P}=0.8$, $0.5$, $0.3$, $0.1$, respectively; The changes of
heat capacity for different $\bar{\phi}_{P}$ in figure~\ref{Cvb}~(a)
with $\chi_r$ is shown in figure~\ref{Cvb}~(b). The squares,
triangles, diamonds and hexagons denote the case of
$\bar{\phi}_{P}=0.8$, $0.5$, $0.3$, $0.1$, respectively. \label{Cvb}}
\end{figure}

In order to demonstrate the effects of the chain length $N$ and polymer
concentration $\bar{\phi}_{P}$ on aggregation of micelles, the
variations of the average volume fraction of stickers $\bar{\phi
}_\textrm{co}^\textrm{s}$ and the lattice site numbers $N_\textrm{co}^\textrm{ls}$ at the
micellar cores ($\phi _\textrm{co}^\textrm{s}\geqslant0.5$) with $\chi_r$ (the $\chi$
deviation from micellar boundary) in various polymer concentrations,
for $N=26$ and $N=34$, are presented in figure~\ref{Cva}~(a) and
figure~\ref{Cvb}~(a), respectively. For $N=26$, at
$\bar{\phi}_{P}=0.1$, $\bar{\phi }_\textrm{co}^\textrm{s}$, as well as the
corresponding $N_\textrm{co}^\textrm{ls}$,  smoothly rises with $\chi_r$ and then
remains constant. When $\chi_r\geqslant0.5$, $N_\textrm{co}^\textrm{ls}$ does not
change with $\chi_r$, and the aggregation degree of micelles
 strengthens from the increase in $\bar{\phi }_\textrm{co}^\textrm{s}$.
When $\bar{\phi}_{P}$ is increased the change of  $\bar{\phi
}_\textrm{co}^\textrm{s}$ with $\chi_r$ is not monotonous. At
$\bar{\phi}_{P}=0.3$ and $\bar{\phi}_{P}=0.5$, when $\chi_r$ is
increased, $\bar{\phi }_\textrm{co}^\textrm{s}$ firstly rises, and then a
$\bar{\phi }_\textrm{co}^\textrm{s}$-lower region occurs in the range of $\chi_r$,
and $\bar{\phi }_\textrm{co}^\textrm{s}$ finally tends to be constant. The
corresponding $N_\textrm{co}^\textrm{ls}$ firstly rises, and then a jump occurs at
the onset of the above lower region. When $\bar{\phi}_{P}=0.8$,
$\bar{\phi }_\textrm{co}^\textrm{s}$ always goes up with $\chi_r$, going with
the slight surge of $N_\textrm{co}^\textrm{ls}$. It is seen that, at intermediate
concentrations, when $\chi_r\geqslant1.1$, micelles dissolve and remicellize,
which is demonstrated by a decrease of the average volume fraction
of stickers at micellar core with increasing from $\chi_r=1.3$ to
$\chi_r=1.4$ (see figure~\ref{dist}). This can accelerate the further
aggregation of micelles. At high concentrations, the behavior of
micellar dissolution and remicellization is restrained. Only a few
micelles dissolve to strengthen the aggregation degree of micelles.
It is shown that the way of an increase in aggregation degree of
micelles depends on polymer concentration. It is noted that when
$\chi_r\geqslant0.6$, $\bar{\phi }_\textrm{co}^\textrm{s}$ at fixed $\chi_r$ decreases
with an increase in $\bar{\phi}_{P}$ for $N=26$.

For $N=34$ (see figure \ref{Cvb}~(a)), when polymer concentration
($\bar{\phi}_{P}=0.1$ and $0.3$) is not high, the tendencies of
$\bar{\phi }_\textrm{co}^\textrm{s}$ and $N_\textrm{co}^\textrm{ls}$ to $\chi_r$ are similar to
those of $N=26$. At intermediate and high polymer concentrations,
they are different from those of $N=26$. When $\bar{\phi}_{P}=0.5$,
$\bar{\phi }_\textrm{co}^\textrm{s}$ and $N_\textrm{co}^\textrm{ls}$ always smoothly increase
with $\chi_r$. At $\bar{\phi}_{P}=0.8$, $\bar{\phi }_\textrm{co}^\textrm{s}$
always smoothly increase with $\chi_r$, but $N_\textrm{co}^\textrm{ls}$ goes
down slowly with $\chi_r$. It is demonstrated that micelles
almost do not dissolve at $\bar{\phi}_{P}=0.5$. Consequently, the
micellar further aggregation is restrained in a way near the
micellar boundary. $\bar{\phi }_{\mathrm{co}}^{\mathrm{s}}$ at $\bar{\phi}_{_P}=0.5$ is larger than that of
           $\bar{\phi}_{_P}=0.8$  until $\chi_r\geqslant1.5$,
which is larger from the case of $N=26$. It is demonstrated that, at intermediate and high concentrations, the
further aggregation of micelles is markedly affected by the increase
in $N$.

 The heat capacity is an important thermodynamic signature to
test the occurrence of a phase transition in a system.
 The shape of specific
heat peak may also be a characteristic of
transition. \cite{Doug2006,Han2012}. In this work, the heat capacity
 per site of amphiphilic  symmetric ABA triblock copolymers is
expressed as (in the unit of $k_\textrm{B}$):
\begin{eqnarray} C_{V} =\left(\frac{\partial {U}}{\partial
{T}}\right)_{N_{{L}},{n}_{{P}}}
=\frac{1}{N_{{L}}}\chi ^{2}\frac{\partial }{\partial {\chi
}}\left( \sum_{r}\phi _\textrm{st}^{2}(r)\right)
\label{scf5-2}.
\end{eqnarray}%
The $C_{V}(\chi_r)$ curves for the unimer-micelle transition in
various  $\bar{\phi}_{{P}}$ at $N=26$ and $N=34$ are shown in
figure~\ref{Cva}~(b) and figure~\ref{Cvb}~(b), respectively. For
unimer-micelle transition, an asymmetric specific heat peak appears.
For $N=26$, when $\bar{\phi}_{{P}}=0.1$, there is only a peak on
$C_{V}(\chi_r)$ curves. when $\bar{\phi}_{{P}}$ is increased, a
primary and a secondary peaks, are observed as shown in
figure~\ref{Cva} (at intermediate and high concentrations). When
$\bar{\phi}_{{P}}=0.3$ and $0.5$, the primary peak is higher than
the corresponding secondary peak. When $\bar{\phi}_{{P}}=0.8$, a
primary and two secondary peaks occur, and one of them is nearly as
high as the primary peak. The primary and secondary peaks tend to be
similar. The occurrence of the secondary peak is according to the
saltation of the $\chi_r$ curves of the average volume fraction of
stickers $\bar{\phi}_\textrm{co}^\textrm{s}$ and the lattice site numbers
$N_\textrm{co}^\textrm{ls}$ at the micellar cores. Ignoring the secondary peak,
the specific heat peak becomes broad with increasing
$\bar{\phi}_{{P}}$. For $N=34$, when polymer concentration is low,
the peak is narrow and similar to the corresponding case of $N=26$.
When $\bar{\phi}_{{P}}$ is increased, the peak shape changes and
the peak also becomes broad. With increasing $\bar{\phi}_{{P}}$,
the maximum of $C_{V}$ shifts to a big $\chi_r$ and the curves of
$C_{V}(\chi_r)$ tend to be not smooth, thus the broad peak seems to
be a primary peak. For a long chain, the peak shape changes
markedly at intermediate and high concentrations.

\begin{figure}[!t]
\centering
\includegraphics[width=0.8\textwidth]{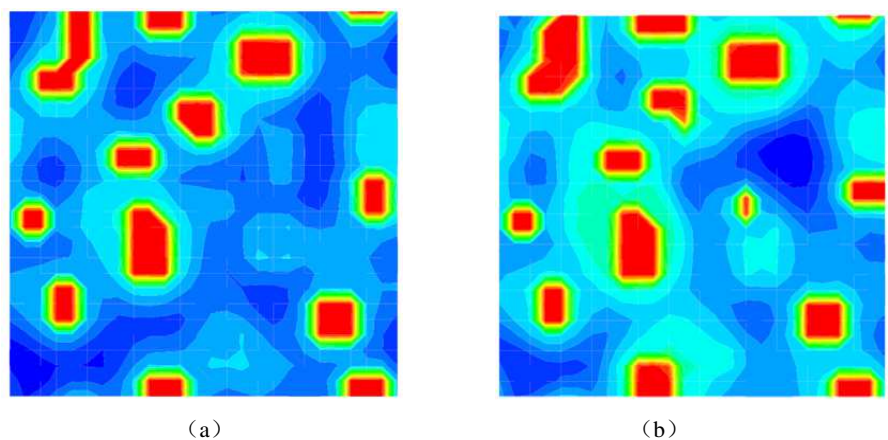}
\caption{(Color online) The cross sections of the system  are presented in
figure~\ref{cross}. Figure~\ref{cross}~(a) and (b), which demonstrate
the changes concerned with remicelliztion, correspond to the cases
of $\chi_r=1.3$ and $\chi_r=1.4$, respectively, for
$\bar{\phi}_{P}=0.3 $ and $N=26$. \label{cross}}
\end{figure}

When temperature drops to a certain extent, micelles appear, and with
a further decrease in temperature the aggregation degree  of  micellar
cores markedly strengthens. It is indicated above that the
temperature-dependent aggregation behavior of micelles depends
on polymer concentration and the chain length. The micellization of
hydrophobic end blocks of triblock copolymer can be considered in
the following ways. One is that both end blocks of each individual
molecule could be incorporated into the same core, the other is that
the two hydrophobic ends of triblock copolymer could be incorporated
into two adjacent micelles. For a short chain, at low
concentration, the first micellization way is dominant. When polymer
concentration is increased, the possibility of the two hydrophobic
ends of triblock copolymer to be incorporated into two adjacent
micelles will rise markedly. At the same time, the  aggregation
degree of micelles on micellar boundary tends to decrease.
Therefore, at intermediate concentrations, the further micellization
of triblock copolymers is delayed for a while due to the
correlations from chain connection among micelles. When temperature
is decreased to some extent, many micelles will be dissolved, and
then remicellize. On remicellization, the distribution width of
the volume fraction of stickers at micellar cores increases
which is different from the general variation of distribution for volume fraction of stickers at micellar cores with $\chi_r$. At the
same time, the relationship among the micelles becomes stronger, as
shown in figure \ref{cross}. At high concentrations, the
correlations among micelles strengthens compared with intermediate
concentrations, only a few micelles dissolve and some new
micelles form, which is demonstrated by keeping the
tendency of $\bar{\phi }_\textrm{st}(\chi_r)$ to increase and the surge of
$N_{ls}(\chi_r)$ with $\chi_r$. For a long chain, when concentration
is low, the case is similar to that of a short chain. When the
polymer concentration is increased, the chain length effect emerges. The
two hydrophobic ends of triblock copolymer will be almost absolutely
incorporated into two adjacent micelles or small
aggregates ($\phi _{\mathrm{co}}^{\mathrm{s}}<0.5$ in figure~\ref{dist}). When $\bar{\phi}_{{P}}=0.5$, quite a few small
aggregates form. The increase in aggregation degree of micelles
is caused by dissolution of small aggregates, and thus the
corresponding $N_\textrm{co}^\textrm{ls}$ does not surge. At high concentrations,
micelles form easily, and the quantity of small aggregates decreases
notably. Thus, the further aggregation of micelles results from the
dissolution of a few micelles. In summary, at intermediate
concentrations for a short chain, the increase in aggregation
degree of micelles is by its dissolution and remicellization. When polymer concentration or chain length is
increased, the way of dissolution and remicellization is restrained,
and thus the temperature-dependent aggregation behavior of micelles
is changed.

In the end of paper, the validity of self-consistent field theory for
the above results should be illuminated. Capture can be an essential feature for
the accounted effect of polymer concentration $\bar{\phi}_{{P}}$
and chain length $N$. The chain accounted in
the work should belong to gaussian chain. Using a self-consistent
field lattice model, the phase diagram of coil-coil diblock
copolymers for $N=20$ in the three dimension space \cite{Chen2006} is
consistent with the Matsen-Schick phase diagram. \cite{Mats1994}, and
the results for the solution of homopolymer length $N=30$ in a two
dimensional square lattice is also in reasonable agreement with the
theoretical prediction \cite{Han2010}. Furthermore, the effect of
relative chain length is also accounted for polymer
blends \cite{Mats1995}. The above effects from chain length $N=26$ and
$N=34$ should be reasonable. SCFT is extensively applied to the study
of the phase behavior of dilute amphiphilic block copolymer
solutions, and the obtained results at $\bar{\phi}_{{P}}=0.1$ have
been proved by experimental observations \cite{Wang2011}. The
specific heat peak for the transition concerned with micelles is
also calculated in physically associating polymer
solutions \cite{Han2010}, and the effect of concentration on specific
heat peak (not shown) is in reasonable agreement with that of the related
system \cite{Dudo1999}. Therefore, the concentration effect accounted for
in the work by self-consistent field theory is reasonably
qualitative.

\clearpage

\section{Conclusion and summary\label{sec4}}

Using the self-consistent field lattice model, polymer concentration
$\bar{\phi}_{{P}}$ and the chain length $N$ (the length ratio of
hydrophobic to hydrophilic blocks remains constant), the effects on the aggregation
behavior of micelles are studied in amphiphilic symmetric ABA
triblock copolymer solutions. When $N$ is increased, at fixed
$\bar{\phi}_{{P}}$, micelles occur at a higher temperature. The
variations of the average volume fraction of stickers $\bar{\phi
}_\textrm{co}^\textrm{s}$ and the lattice site numbers $N_\textrm{co}^\textrm{ls}$ at the micellar
cores with temperature depend on $N$ and $\bar{\phi}_{{P}}$,
which is demonstrated by the change of the specific heat peak. For a
short chain, when $\bar{\phi}_{{P}}$ is increased, firstly a
peak appears on the curve of $C_V$ for the micellar appearance, and
then, in addition to a primary peak, the secondary peak is observed. For
a long chain, in intermediate and high concentration regimes, the
shape of a specific heat peak changes markedly, and it tends to a
broader primary peak, which is explained by the aggregation way
of amphiphilic triblock copolymer. For a short chain, at
intermediate concentrations, the way of two hydrophobic ends of
triblock copolymer to be incorporated into two adjacent micelles is
dominant. Therefore, the aggregation degree of the micelles increases
by its dissolution and remicellization. When
polymer concentration or chain length is increased, the way of
dissolution and remicellization is restrained, and thus the
temperature-dependent aggregation behavior of the micelles changes.

\section*{Acknowledgements} This research is financially supported
by the National Nature Science Foundations of China (11147132) and
the Inner Mongolia municipality (2012MS0112), and the Innovative
Foundation of Inner Mongolia University of Science and Technology
(2011NCL018).

\newpage

\ukrainianpart

\title{Самоузгоджене теоретико-польове моделювання
амфіфільних  триблочних кополімерних розчинів: \\
ефекти концентрації та довжини ланцюга полімерів}
\author{К.-Г. Ган, Й.-Г. Ма}
\address{
Університет науки і технологій Внутрішньої Монголії,  Баоту 014010, Китай}

\makeukrtitle

\begin{abstract}
\tolerance=3000%
Використовуючи самоузгоджену польову ґраткову модель,  вивчаються ефекти концентрації $\bar{\phi}_{{P}}$
і довжини ланцюга $N$  полімера (при фіксованому відношенні довжин гідрофобних і гідрофільних блоків)
на температуро-залежну поведінку міцел в амфіфільних симетричних  ABA триблочних кополімерних розчинах.
Якщо довжина ланцюга зростає, при фіксованому  $\bar{\phi}_{{P}}$, міцели утворюються при вищій температурі.
Зміна середньої об'ємної долі стикерів  $\bar{\phi }_\textrm{co}^\textrm{s}$ та числа вузлів ґратки
$N_\textrm{co}^\textrm{ls}$ з температурою при  міцелярних корах залежать від  $N$ і $\bar{\phi}_{{P}}$, що вказує на залежність
агрегації міцел від  $N$ і $\bar{\phi}_{{P}}$.
Крім того, якщо  $\bar{\phi}_{{P}}$ зростає, спочатку пік виникає на кривій питомої теплоємності  $C_V$ для
переходу мономер-міцела, і потім додатково спостарігаються на кривій $C_V$ основний пік, вторинний пік,
які є результатом реміцелізації. Для довгого ланцюга при режимах проміжних і високих концентрацій значно змінюється форма піку питомої теплоємності, і пік прямує до ширшого  піку. Накінець, агрегаційна поведінка міцел пояснюється способом агрегації амфіфільного триблочного кополімера. Отримані результати є корисними для розуміння процесу агрегації міцел.

\keywords міцела, самоузгоджене поле, амфіфільний кополімер
\end{abstract}

 \end{document}